# A HELICAL COOLING CHANNEL SYSTEM FOR MUON COLLIDERS[*]

K. Yonehara[#], Fermilab, Batavia, IL 60510, U.S.A.
R. P. Johnson, M. Neubauer, Muons, Inc., Batavia, IL 60510, U.S.A.
Y. S. Derbenev, Jlab, Newport News, VA 23606, U.S.A.

*Abstract*

Fast muon beam six dimensional (6D) phase space cooling is essential for muon colliders. The Helical Cooling Channel (HCC) uses hydrogen-pressurized RF cavities imbedded in a magnet system with solenoid, helical dipole, and helical quadrupole components that provide the continuous dispersion needed for emittance exchange and effective 6D beam cooling. A series of HCC segments, each with sequentially smaller aperture, higher magnetic field, and higher RF frequency to match the beam size as it is cooled, has been optimized by numerical simulation to achieve a factor of $10^5$ emittance reduction in a 300 m long channel with only a 40% loss of beam. Conceptual designs of the hardware required for this HCC system and the status of the RF studies and HTS helical solenoid magnet prototypes are described.

## INTRODUCTION

Because a muon has 210 times the mass of the electron, its synchrotron radiation is enormously reduced and it can be accelerated and stored in a circular ring. Therefore, the size of the accelerator complex for a muon collider can be much smaller than such as a CLIC type machine with the same Center of Mass (CM) energy. Recently, the Muon Accelerator Program (MAP) has been established to investigate how to generate a high-energy low-emittance muon beam [1]. A goal of the MAP is accelerating and colliding muons within their lifetime, which is 2.2 μsec at rest and in 31 msec at 3 TeV CM. However, the initial phase space of muons that are generated via pion decay is too large to fit in a conventional accelerator system, and fast 6D phase space cooling is required.

Ionization cooling has been proposed for this purpose, where muons pass through a low-Z absorber, losing energy as they ionize the absorber material. The lost longitudinal energy is regained by RF cavities while the transverse energy shrinks until it comes into equilibrium with heating from multiple Coulomb scattering.. Only the transverse phase space becomes smaller by this ionization cooling process, so that 6D phase space ionization cooling requires emittance exchange, a straightforward feature of the HCC [2]. A HCC consists of solenoid, helical dipole, and helical quadrupole magnetic field components that enclose hydrogen gas filled RF cavities. The helical magnet generates a continuous dispersion such that the larger energy loss for higher momentum muons due to their longer path length reduces the energy spread.

However, the transverse emittance is increased in this process so that the emittance is exchanged and although only ionization cooling is only transverse, effective 6D cooling has been shown to work as the theory predicts by numerical simulations [3, 4].

Current HCC cooling simulation efforts aim to optimize the lattice parameters based on designs of realistic magnets and RF structures. This document summarizes the recent simulations used to guide to the engineering design and the present status of those designs.

## SIMULATION

The most recent 6D phase space cooling numerical simulations for an optimized HCC system using G4beamline [5] is shown in Fig. 1. Here the phase space evolution starts from top right in this Fernow-Neuffer plot of average transverse emittance versus longitudinal emittance. The acceptable momentum spread and beam size of the first HCC segment are $\Delta p/p = \pm 22\%$ and 150 mm, respectively. Those values are larger than the beam at the end of Study2A neutrino factory design frontend channel (blue point in the plot).

The transverse and longitudinal phase spaces are cooled down in a series of HCCs in which the magnetic field strengths are increased as a function of channel length (Table 1). The stronger magnetic field confines a muon beam in smaller size, i.e. the beta function goes down from 0.27 m to 0.09 m. The transmission efficiency is 60% (25% loss due to muon decay and 15% due to matching) in 300 m channel length. The obtained merit factor (transmission efficiency × Initial 6D emittance/Final 6D emittance) is over $10^5$.

A continuous dense hydrogen gas filled RF cavity is incorporated into the helical magnet. An ideal pillbox cavity with $1.3 \times 10^{-2}$ g/cm$^3$ hydrogen gas (160 atm at room temperature) and 60-μm Be RF windows are modeled in the simulation. The RF electric field is parallel to the solenoid field axis. There are 20 cavities per helical period. The RF amplitude is 28 MV/m (reference RF phase = 160 degrees) to recover the ionization energy loss in the channel. Because the muon beam size gets shorter as it gets cooled, the RF cavity frequency can be increased in the later HCC sections. The RF frequency starts at 0.325 GHz and increases to 1.3 GHz in the last section.

Table 1 summarizes the lattice parameters and emittances at the end of each HCC section except for section "*1*" that shows the acceptance of the first HCC. The HCC lattice is designed for a 200 MeV/c left-handed helicity (counter-clockwise rotation) for a positive muon. The helicity and sign of fields must be carefully chosen for the opposite sign of muon beam.

*Work supported by DOE STTR grant DE-FG02-07EER84825 & FRA under DOE contract DE-AC02-07CH11359
#yonehara@fnal.gov

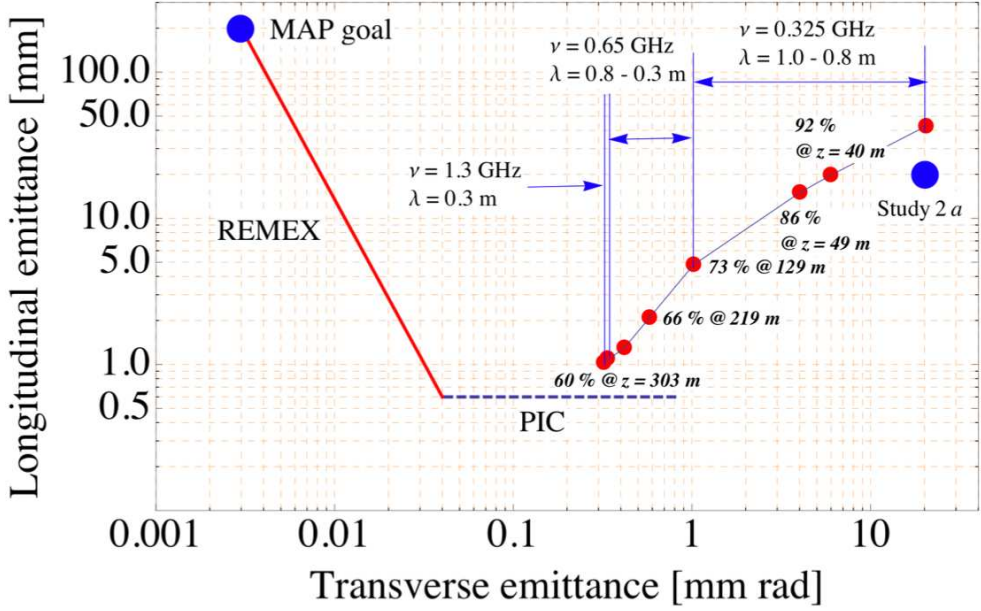

Figure 1: Transverse – Longitudinal phase space evolution in a series of HCCs. PIC and REMEX are Parametric Ionization Cooling and Reverse EMittance EXchange for final muon beam cooling [6].

| | $z$ | $b$ | $b'$ | $b_z$ | $\lambda$ | $N$ | $\varepsilon_T$ | $\varepsilon_L$ | $\varepsilon_{6D}$ | $\varepsilon$ |
|---|---|---|---|---|---|---|---|---|---|---|
| unit | m | T | T/m | T | m | GHz | mm rad | mm | mm$^3$ | Transmission |
| 1 | 0 | | | | | | 20.4 | 42.8 | 12900 | 1.00 |
| 2 | 40 | 1.3 | -0.5 | -4.2 | 1.0 | 0.325 | 5.97 | 19.7 | 415.9 | 0.92 |
| 3 | 49 | 1.4 | -0.6 | -4.8 | 0.9 | 0.325 | 4.01 | 15.0 | 10.8 | 0.86 |
| 4 | 129 | 1.7 | -0.8 | -5.2 | 0.8 | 0.325 | 1.02 | 4.8 | 3.2 | 0.73 |
| 5 | 219 | 2.6 | -2.0 | -8.5 | 0.5 | 0.65 | 0.58 | 2.1 | 2.0 | 0.66 |
| 6 | 243 | 3.2 | -3.1 | -9.8 | 0.4 | 0.65 | 0.42 | 1.3 | 0.14 | 0.64 |
| 7 | 273 | 4.3 | -5.6 | -14.1 | 0.3 | 0.65 | 0.32 | 1.0 | 0.08 | 0.62 |
| 8 | 303 | 4.3 | -5.6 | -14.1 | 0.3 | 1.3 | 0.34 | 1.1 | 0.07 | 0.60 |

Table 1: Field parameter and beam parameter in a series of HCCs

| | Radius of HS coil center position | $\lambda$ | $b_{solenoid}$ | HS coil inner radius | HS coil outer radius | # of coils / Coil length | Current density | Length |
|---|---|---|---|---|---|---|---|---|
| unit | m | m | T | m | m | m | A/mm$^2$ | m |
| Upstream match section | 0 $\rightarrow$ 0.28 | 1.0 | 0.55 | 0.35 | 0.40 | 20/0.025 | -220.0 $\rightarrow$ -194.0 | 5.5 |
| Cool section 2 | 0.28 | 1.0 | 0.55 | 0.35 | 0.40 | 20/0.025 | -194.0 | |
| Cool section 6 | 0.16 | 0.4 | 6.73 | 0.18 | 0.28 | 20/0.010 | -332.9 | |

Table 2: HS coil parameters for Upstream matching and Cooling section "2" and "6" in Table 1

## BEAM ELEMENTS

### Helical solenoid magnet

The Helical Solenoid (HS) coil [7] has been investigated to generate the optimum helical field structure as shown in Table 1. The helical magnetic field is tuned by changing the coil geometry. For instance, the coil position (relative displacement from a neighbor coil) primarily changes the helical dipole field strength and the coil radius changes the helical quadrupole field strength. The HS coil geometry and current density parameters for Cool sections *2* and *6* in Table 1 are shown as examples in Table 2. The design current densities in sections *2* and *6*"are well below the critical current of *Nb-Ti* and *Nb$_3$-Sn*, respectively.

We also designed a matching section in which an initial pure solenoid focused coaxial beam is transported into the HCC section *2*, by adiabatically applying introducing helical dipole and helical quadrupole magnetic fields. Such a field is generated by slowly increasing the radius of the HS coil center position. Betatron motion and the phase and dispersion in the matching section are tuned by adjusting the field strength. This can be done by linearly ramping up the current density of the HS coils in a moderately long matching section (the adiabatic condition is preserved when the matching length is more than 3 betatron wavelengths).

A 4-coil HS segment has been made and successfully tested [8] using surplus SSC superconductor. The design, construction, and testing of the high field HS coil for the final HCC section, which is the technically the most crucial part, is underway [9]. High temperature superconductor (HTS) YBCO operating at low temperature will be used for the final HCC sections *7* and *8* in Table 1.

*Pressurized RF cavity*

A hydrogen pressurized RF test cavity has been made and tested in various conditions [10-12]. Only a beam test remains; the cavity Q may be degraded by beam-induced plasma in the dense hydrogen gas. This issue can be solved by adding an electro-negative dopant gas in hydrogen gas. The beam test is scheduled in 2010 [13].

Various RF cavities have been studied [14]. A dielectric loaded RF cavity has been designed with realistic parameters [15, 16]. The transverse size of the dielectric-loaded RF cavity can be smaller than the pillbox. It reduces the size of cavity more than a half transverse size of the pillbox one. However, the enormous amount of RF energy will be dissipated in a dielectric material if its loss tangent is more than $10^{-3}$. Current state of the art produces much lower loss tangent material, e.g. WESGO Alumina Ceramics has a loss tangent ~ $10^{-4}$ at room temperature and < $10^{-4}$ in cryogenic temperature. The estimated RF energy dissipation in this type of material is still high [16]. Sapphire shows much better loss tangent, i.e. it is $10^{-5}$ at room temperature and reaches $10^{-7}$ in cryogenic temperature. The RF power dissipation issue is significantly smaller in the higher frequency RF cavities with their smaller volume.

## INTEGRATING RF INTO THE HS

Integrating RF cavities into the HS magnet is the final design issue. A larger inner diameter HS coil makes the lower helical quadrupole component lower. The field reduction can be compensated by adding a quadrupole correction magnet [17]. More elegantly, additional quadrupole components can be generated by displacing the HS coil center position outward from the magnet center. With the optimized coil center position, the consequent coil size is large enough to incorporate RF cavities in HCC sections *2* and *6* shown in Table 2 even without dielectric inserts.

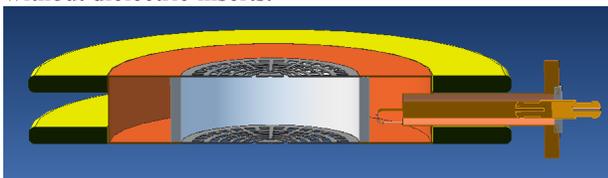

Fig. 2: HCC Cell concept. HTS coils (yellow), dielectric (orange), Be windows (gray), RF coupler (brown).

Figure 2 shows one HCC cell concept. Figure 3 shows a HCC wavelength made of ~20 such cells. Figure 4 shows the general scheme of integrating RF cavities into the helical magnet. In this design, the pressure shell is located outside the HS coil. If the HTS critical current is high enough at 30 K operation we can remove any thermal gap between the RF cavity and the HS coil. This structure would be preferable for the final HCC section.

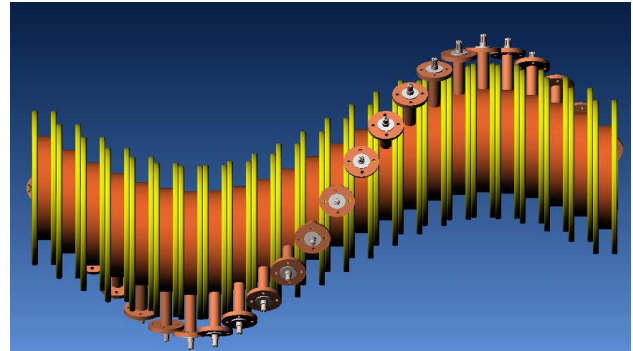

Fig. 3: Schematic of integrated HS coils and RF cavities.

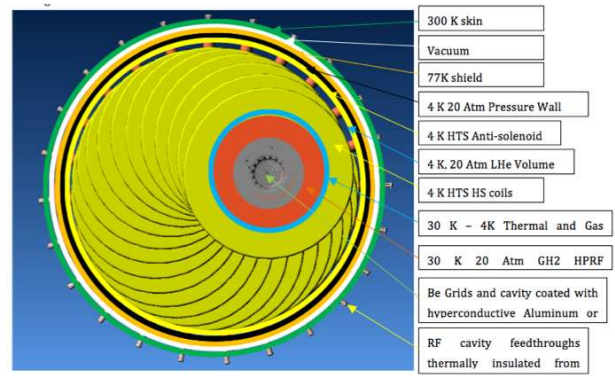

Fig. 4: Conceptual picture of cryomodule components to contain magnet coils and RF cavities.

## REFERENCES

[1] MAP proposal, https://mctf.fnal.gov/mapproposal.pdf/download
[2] Ya. S. Derbenev and R.P. Johnson, PRSTAB **8**, 0412002 (2005).
[3] K. Yonehara et al., PAC'05, TPPP052.
[4] K. Yonehara and V. Balbekov, EPAC'08, THPC110.
[5] T.J. Roberts et al., EPAC'08, WEPP120.
[6] A. Afanasev et al., IPAC'10, MOPEA042.
[7] V. Kashikhin et al., EPAC'08, WEPD015.
[8] M. Lamm et al., PAC'09, MO6PFP059.
[9] A. Zlobin et al., IPAC'10, MOPEB054.
[10] K. Yonehara et al., IPAC'10, WEPE069.
[11] K. Yonehara et al., PAC'09, TU5PFP020.
[12] P. Hanlet et al., EPAC'06, TUPCH147.
[13] M. Chung et al., IPAC'10, WEPE066.
[14] K. Yonehara et al., PAC'09, TU5PFP021.
[15] M.A.C. Cummings et al., IPAC'10, THPEA047.
[16] V. Kashikhin et al., IPAC'10, THPEA047.
[17] S. Kahn et al., IPAC'10, WEPE032.